\documentclass[twocolumn,showpacs,preprintnumbers,amsmath,amssymb]{revtex4}
\usepackage{mathrsfs}%====================================
\usepackage{amssymb}
\usepackage{graphicx}% Include figure files
\usepackage{dcolumn}% Align table columns on decimal point
\usepackage{bm}% bold math
\usepackage{textcomp}
\usepackage{ulem}
\usepackage{multirow}

\newcommand\ket[1]{\ensuremath{|#1\rangle}}
\newcommand\bra[1]{\ensuremath{\langle#1|}}

\newcommand\meanv[1]{\langle#1\rangle}

\begin{document}
\title{Full  Implementation of 4-intensity Protocol for Measurement-Device-Independent Quantum Key Distribution over Asymmetric Channel}
%\title{Measurement-Device-Independent Quantum Key Distribution with High Key Rate: Full Application of 4-Intensity Protocol over Asymmetric Channel}
\author{ Xiao-Long Hu$ ^{1}$, Yuan Cao$ ^{2,3}$, Zong-Wen Yu$ ^{1,5}$,
and Xiang-Bin Wang$ ^{1,4,6\footnote{Email
Address: xbwang@mail.tsinghua.edu.cn}\footnote{Also a member of Center for Atomic and Molecular Nanosciences at Tsinghua University}}$}

\affiliation{ \centerline{$^{1}$State Key Laboratory of Low
Dimensional Quantum Physics, Department of Physics,} \centerline{Tsinghua University, Beijing 100084,
People¡¯s Republic of China}
\centerline{$^{2}$National Laboratory for Physical Sciences at the Microscale and Department of Modern Physics,}
\centerline{University of Science and Technology of China, Hefei 230026, China}
\centerline{$^{3}$CAS Center for Exellence and Synergetic Innovation Center in Quantum Information and Quantum Physics,}
\centerline{University of Science and Technology of China, Shanghai 201315, China}
\centerline{$^{4}$ Synergetic Innovation Center of Quantum Information and Quantum Physics, University of Science and Technology of China}
\centerline{  Hefei, Anhui 230026, China
 }
\centerline{$^{5}$Data Communication Science and Technology Research Institute, Beijing 100191, China}
\centerline{$^{6}$ Jinan Institute of Quantum technology, SAICT, Jinan 250101,
People¡¯s Republic of China}}
%%%%%%%%%%%%%%%%%%%%%%%%%%%%%%%%%%%%%%%%%%%%%%%%%%%%%%%%%%%%%%%%%%%
%%%%%%%%%%%%%%%%%%%%%%%%%%%%%%%%%%%%%%%%%%%%%%%%%%%%%%%%%%%%%%%%%%%
%%%%%%%%%%%%%%%%%%%%%%%%% Abstract %%%%%%%%%%%%%%%%%%%%%%%%%%%%%%%%
\begin{abstract}
%We study the measurement-device-independent quantum key distribution (MDIQKD) in the free space. The biggest challenge in the free-space MDIQKD is that the unstable and asymmetric quantum channel decreases the key rate quite a lot. We present a loss-compensation method with monitoring the channel loss. This method can decrease the error rate caused by the free-space channel and then increase the secure key rate. It will improve the implementation of ultra-long-distance MDIQKD with satellite in space.
We study the optimization of full implementation of the four-intensity decoy-state Measurement-Device-Independent Quantum Key Distribution (MDIQKD) over  asymmetric and unstable quantum channel.  As was requested by the original 4-intensity protocol, Alice and Bob choose different four intensities separately (i.e., 7 intensities for both sides including one vacuum and 6 non-vacuum). We make the optimization with 12 independent parameters taking both the global optimization for the independent parameters and the joint constraints through employing  a gradient optimization method.
Compared with partial implementation, our full implementation of 4-intensity improves the key rate by 1 to tens of times in typical experimental conditions.
In addition, we present a loss-compensation method with monitoring the channel loss. The numerical simulation shows that the method can produce high key rate for both the asymmetric channel and the unstable channel.

\end{abstract}

%%%%%%%%%%%%%%%%%%%%%%%%%%%%%%%%%%%%%%%%%%%%%%%%%%%%%%%%%%%%%%%%%%%
%%%%%%%%%%%%%%%%%%%%%%%%%%%%%%%%%%%%%%%%%%%%%%%%%%%%%%%%%%%%%%%%%%%
%%%%%%%%%%%%%%%%%%%%%%%%%%%%%%%%%%%%%%%%%%%%%%%%%%%%%%%%%%%%%%%%%%%

\maketitle

%%%%%%%%%%%%%%%%%%%%%%%%%%%%%%%%%%%%%%%%%%%%%%%%%%%%%%%%%%%%%%%%%%%
%%%%%%%%%%%%%%%%%%%%%%%%%%%%%%%%%%%%%%%%%%%%%%%%%%%%%%%%%%%%%%%%%%%
%%%%%%%%%%%%%%%%%%%%%%%%%%%%%%%%%%%%%%%%%%%%%%%%%%%%%%%%%%%%%%%%%%%
%%%%%%%% Introducation & Motivation %%%%%%%%%%%%%%%%%%%%%%%%%%%%%%%

\section{Introduction}\label{SecIntro}
Quantum key distribution (QKD) provides the communication users with secure keys to encrypt their information.
Bennett and Brassard proposed BB84 protocol~\cite{bennett1984quantum} to realize QKD, but the lack of practical single-photon sources limited the use of origin BB84 protocol.
BB84 protocol with imperfect single-photon sources would suffer from the photon-number-splitting (PNS) attack~\cite{huttner1995quantum,yuen1996quantum,brassard2000limitations}.
This loophole can be fixed by the decoy-state method~\cite{inamori2007unconditional,hwang2003quantum,wang2005beating,lo2005decoy}.
With the decoy-state method, QKD can be used in the practical system between users with longer distace~\cite{wang2008experimental,xu2009experimental,zhang2015approaching}.
After that, measurement-device-independent QKD (MDIQKD) was proposed to avoid any loophole from the imperfect detection devices~\cite{lydersen2010hacking,gerhardt2011full,braunstein2012side,lo2012measurement}. Combined with decoy-state method, MDIQKD can also avoid the loophole from the imperfect single-photon sources~\cite{lo2012measurement,wang2013three}.
Nowadays,  the decoy-state MDIQKD has become the mainstream of the studies of quantum key distribution both theoretically~\cite{wang2013three,wang2013efficient,curty2014finite,yu2013three,zhou2014tightened,wang2014simulating,xu2014protocol,yu2015statistical,zhou2016making,wang2016realizing,jiang2016measurement,jiang2017measurement,zhou2017obtaining,hu2017practical} and experimentally~\cite{rubenok2013real,chan2014modeling,liu2013experimental,da2013proof,tang2014experimental,tang2014measurement,wang2015phase,wang2017measurement,pirandola2015high}.
The maximum distance of MDIQKD has been experimentally increased to 404 kilometers~\cite{yin2016measurement}.

In the scheme of decoy-state MDIQKD, at each time the user Alice (Bob) randomly chooses her (his) basis, bit value and intensity to send a pulse in a corresponding state, e.g. BB84 state in a polarization-coding MDIQKD, to an untrusted third party (UTP) Charlie.
Charlie performs a collective measurement on each pulse pair and announces the measurement result in the public channel.
After all pulses are sent, Alice and Bob announce the bases and intensities they use. Based on all announcement, Alice and Bob can calculate the yield and the error rate of single-photon pulse pairs, and then distill the secure key.

For the practical applications, the asymmetric and unstable channels are common cases both in fiber and free space.
For example, when we consider the quantum network, due to the different geographical locations of users, the channel losses can be largely different.
And if we want implement MDIQKD in free space, the channels are always asymmetric and unstable too, due to the atmospheric turbulence or moving sites (such as the satellite).
Although the security of MDIQKD doesn't make any assumption to the channel, the unstable and/or asymmetric quantum channel decreases the key rate quite a lot.
Therefore, directly applying the origin decoy-state MDIQKD in the asymmetric and unstable channel doesn't give a good performance.
Here, we propose full implementation of four-intensity decoy-state MDIQKD protocol to largely increase the key rate in asymmetric channels than previous protocol~\cite{xu2013pratical,wang2018enabling}.
Moreover, a loss-compensation method is presented with optimization to increase the performance of MDIQKD in unstable channels.
%The goal of QKD is to be faster and further. One way to realize ultra-long-distance QKD is to implement MDIQKD with satellite in space, due to the near zero loss in space.
%The quantum channel between the satellite and the ground base station is the atmosphere, which is unstable, instead of the stable quantum channel in the fiber.

This paper is arranged as follows.
We give full implementation of four-intensity decoy-state MDIQKD protocol and the method to optimize the source parameters in Sec.~\ref{sec:4-int-MDI}.
Then we introduce our loss-compensation method and the simulation method in Sec.~\ref{sec:method}.
We show the numerical simulation results with parameters fully optimized for both the asymmetric channel and the unstable channel in Sec.~\ref{sec:numerical}.
We summarize in Sec.~\ref{sec:Conclusion}.

\section{full implementation of four-intensity decoy-state MDIQKD protocol}\label{sec:4-int-MDI}
As was stated in the original four-intensity decoy-state method in~\cite{zhou2016making}, Alice and Bob each uses 4 different intensities, including one vacuum. This means in general, there are 7 different intensities for both sides with 6 independent parameters for non-vacuum intensities. Together with the frequencies of using each intensities, there are 12 independent parameters in general in the original protocol. Also, the original 4-intensity protocol suggests using the joint constraints~\cite{yu2015statistical}. A full implementation means doing optimization among all those 12 parameters with joint constraints fully.

Explicitly, in the original four-intensity decoy-state method in~\cite{zhou2016making}, Alice (Bob) uses a source $z_A$ ($z_B$) that only emits photons in the $Z$ basis, two sources $x_A$ and $y_A$ ($x_B$ and $y_B$) that only emit photons in the $X$ basis and a vacuum source $o_A$ ($o_B$) that only emits vacuum pulses.
At each time, Alice (Bob) randomly chooses a source in the four sources above to send a pulse, with probability $p_{al_A},l=z,x,y,o$ ($p_{br_B},r=z,x,y,o$).
So we call it ``four-intensity protocol''.
In photon number space, the density matrices of the pulses from these sources can be written as
\begin{equation}\label{equ:rhol}
    \rho_{l_A} = \sum_{k=0}^{\infty} a_{lk} \ket{k}\bra{k},\ l=x,y,z
\end{equation}
and
\begin{equation}\label{equ:rhor}
    \rho_{r_B} = \sum_{k=0}^{\infty} b_{rk} \ket{k}\bra{k},\ r=x,y,z.
\end{equation}
We assume that the states above satisfy these conditions:
\begin{equation}\label{equ:abcondition}
    \frac{a_{yk}}{a_{xk}} \ge \frac{a_{y2}}{a_{x2}} \ge \frac{a_{y1}}{a_{x1}},\  \frac{b_{yk}}{b_{xk}} \ge \frac{b_{y2}}{b_{x2}} \ge \frac{b_{y1}}{b_{x1}}
\end{equation}
for $k>2$, so that the decoy-state results can apply. Familiar sources used in practice, such as weak-coherent-state sources and heralded single-photon sources out of the parametric-down conversion, satisfy the conditions above.

In the following, we will omit the subscript $A$ and $B$ in $l_A$ and $r_B$ if it doesn't cause confusion.

\subsection{Asymptotic case}
We define the yield, the error yield and the error rate as follow.
Consider a pulse pair set $\mathcal{C}$, which contains $N_\mathcal{C}$ pulse pairs totally.
These pairs cause $M_\mathcal{C}$ effective counts and $W_\mathcal{C}$ error counts.
In this case, the yield $S_\mathcal{C}=M_\mathcal{C}/N_\mathcal{C}$, the error yield $T_\mathcal{C}=W_\mathcal{C}/N_\mathcal{C}$ and the error rate $E_\mathcal{C}=W_\mathcal{C}/M_\mathcal{C}$.

The main idea of decoy state is that the yield of $\ket{m}\ket{n}$ photon pairs from different source pairs should be the same in the asymptotic case, which means
\begin{equation}\label{equ:smn}
    \meanv{s_{mn}^{lr}}=\meanv{s_{mn}},\ l,r=x,y,z.
\end{equation}
Using Eq.(\ref{equ:smn}) and the convex form of yield of $lr$ source pairs
\begin{equation}\label{equ:Slr}
    \meanv{S_{lr}}=\sum_{m,n=0}^\infty a_{lm} b_{rn} \meanv{s_{mn}},
\end{equation}
we can calculate the lower bound of the yield of single-photon pairs:
\begin{equation}\label{equ:s11}
    \meanv{s_{11}} \ge \underline{\meanv{s_{11}}} = \frac{S_+ - S_- - a_{y1}b_{y2} \mathcal{H}}{a_{x1}a_{y1} (b_{x1}b_{y2} - b_{x2}b_{y1})}
\end{equation}
where
\begin{equation}\label{equ:S+}
    S_+ = a_{y1} b_{y2} \meanv{S_{xx}} + a_{x1} b_{x2} a_{y0} \meanv{S_{oy}} + a_{x1} b_{x2} b_{y0} \meanv{S_{yo}},
\end{equation}
\begin{equation}\label{equ:S-}
    S_- = a_{x1} b_{x2} \meanv{S_{yy}} + a_{x1} b_{x2} a_{y0} b_{y0} \meanv{S_{oo}}
\end{equation}
and
\begin{equation}\label{equ:H}
    \mathcal{H} = a_{x0} \meanv{S_{ox}} + b_{x0} \meanv{S_{xo}} - a_{x0} b_{x0} \meanv{S_{oo}}.
\end{equation}
Eq.(\ref{equ:s11}) holds when
\begin{equation}\label{equ:Ka<Kb}
    K_a = \frac{a_{y1}a_{x2}}{a_{x1}a_{y2}} \le \frac{b_{y1}b_{x2}}{b_{x1}b_{y2}} = K_b.
\end{equation}
In the case of $K_a > K_b$, the lower bound of $s_{11}$ can be calculated with Eqs.(\ref{equ:s11})-(\ref{equ:H}) by making exchange between $a_{xk}$ and $b_{xk}$, and exchange between $a_{yk}$ and $b_{yk}$, for $k=1,2$.

Similarly, we can calculate the upper bound of phase-flip error rate of single-photon pairs:
\begin{equation}\label{equ:e11}
    \meanv{e_{11}^{ph}} \le \overline{\meanv{e_{11}^{ph}}} = \frac{\meanv{T_{xx}} - \mathcal{H}/2}{a_{x1} b_{x1} \underline{\meanv{s_{11}}}}.
\end{equation}

With $\underline{\meanv{s_{11}}}$ and $\overline{\meanv{e_{11}^{ph}}}$, we can calculate the key rate by:
\begin{equation}\label{equ:keyrate}
    R = p_{z_A} p_{z_B} \{a_{z1} b_{z1} \underline{\meanv{s_{11}}} [1-H(\overline{\meanv{e_{11}^{ph}}})] - f S_{zz} H(E_{zz})\}
\end{equation}
where $H(p)=-p \ln p - (1-p) \ln (1-p)$ and $f$ is the correction efficiency.

\subsection{Nonasymptotic case}
In the nonasymptotic regime, we should consider the statistical fluctuation of the observable, e.g. the difference between observed values and mean values. Given a failure probability $\epsilon$, the observed value $S_\mathcal{C}$ of an observable of a set $\mathcal{C}$ and its mean value $\meanv{S_\mathcal{C}}$ satisfy:
\begin{equation}\label{equ:fluc}
    -\Delta_- \le S_\mathcal{C} - \meanv{S_\mathcal{C}} \le \Delta_+.
\end{equation}
If we perform a standard error analysis, $\Delta$ can be given by
\begin{equation}\label{equ:Delta}
    \Delta_- = \Delta_+ = \gamma \sqrt\frac{S_\mathcal{C}}{N_\mathcal{C}}
\end{equation}
where $N_\mathcal{C}$ is the number of elements in set $\mathcal{C}$ and $\gamma=5.3$ given the failure probability $\epsilon=10^{-7}$.

According to the idea of joint constraints of statistical fluctuation~\cite{yu2015statistical}, the set $\mathcal{C}$ can be either all pulse pairs from a source pair, or the combination of all pulse pairs from different source pairs. Taking all these joint constraints into consideration, the bound of $\meanv{s_{11}}$ and $\meanv{e_{11}^{ph}}$ can be calculated tighter.

As a joint term in $\underline{\meanv{s_{11}}}$ and $\overline{\meanv{e_{11}^{ph}}}$, $\mathcal{H}$ should fluctuate jointly in Eq.(\ref{equ:s11}) and Eq.(\ref{equ:e11}), instead of taking the worst case independently~\cite{zhou2016making}. We regard $R$ as a function of $\mathcal{H}$, scan $\mathcal{H}$ in its possible range and take the minimum $R$ as the final key rate:
\begin{equation}\label{equ:RH}
    \tilde{R} = \min_{\mathcal{H}\in[\underline{\mathcal{H}},\overline{\mathcal{H}}]} R(\mathcal{H})
\end{equation}

\subsection{Optimization of source parameters}
In the numerical simulation, we will estimate what values we would observe for the yields and error rates in a certain model and use these values to calculate the key rate.
So we can regard the key rate as a function of source parameters:
\begin{equation}\label{equ:R1}
\begin{split}
    \tilde{R} = \tilde{R}(&\mu_{ax},\mu_{ay},\mu_{az},p_{ax},p_{ay},p_{az},\\
   & \mu_{bx},\mu_{by},\mu_{bz},p_{bx},p_{by},p_{bz}) = \tilde{R}(\vec{x})
\end{split}
\end{equation}
where $\mu_{al}$ and $\mu_{br}$ are the intensities of Alice's and Bob's sources. If we use weak coherent state sources, the relation between the intensity and the photon number distribution is $a_k=e^{-\mu}\frac{\mu^k}{k!}$.

In the calculation of the key rate, we need to take the joint fluctuation and the scan of $\mathcal{H}$ into consideration.
In addition, the number of parameters we need to optimize is large.
Therefore, normal optimization method costs a lot of time.
We should improve the optimization method to get the optimal parameters quickly.
Firstly, we consider the ``gradient'' of the key rate function~\cite{hu2017practical}:
\begin{equation}\label{equ:grak}
    \frac{\Delta\tilde{R}}{\Delta x_k} = \frac{\tilde{R}(x_k+\Delta x_k,x_i)-\tilde{R}(x_k-\Delta x_k,x_i)}{2\Delta x_k}.
\end{equation}
In the case that both $\tilde{R}(x_k + \Delta x_k,x_i)$ and $\tilde{R}(x_k - \Delta x_k,x_i)$ are less than $\tilde{R}(x_k,x_i)$, we set $\frac{\Delta\tilde{R}}{\Delta x_k}=0$. With
\begin{equation}\label{equ:gra}
    \frac{\Delta\tilde{R}}{\Delta \vec{x}} = (\frac{\Delta\tilde{R}}{\Delta x_1}, \cdots, \frac{\Delta\tilde{R}}{\Delta x_{12}})
\end{equation}
we can find the direction that key rate increases the fastest and get close to the optimal parameters quickly.

To avoid the case that the optimal parameters are the local optimal point, which satisfies $\tilde{R}(x_k+\Delta x_k,x_i) = \tilde{R}(x_k,x_i)$ for any $k$, we search the points in the nearby area to see whether there is higher key rate.
Accurately, we calculate the key rate $\tilde{R}(x_k+\delta_k \Delta l);\delta_k=-1,0,1;k=1,\cdots,12$ with a certain $\Delta l$.
If there are some points with higher key rate, we jump to the point with highest key rate in the nearby area and execute the above procedure again.

In our simulation, we found that in most cases, the gradient method brings us to the optimal point. But in some cases, it brings us to the local optimal point.

\section{Loss-compensation method}\label{sec:method}
For the case of unstable channel, according to our MDIQKD protocol, all source parameters should be determined before the QKD process.
Even though we can detect the channel transmittance $\eta$ at any time, we cannot change the source parameters to optimize the key rate at real time.
With the source parameters fixed, there are always some cases that the intensities at the two sides of Charlie's beam splitter deviate a lot, saying that $\mu_A\eta_A$ and $\mu_B\eta_B$ deviate a lot.
These cases will give a quite high error rate that decreases the key rate a lot.

Consider the case that $\mu_A\eta_A>\mu_B\eta_B$.
If we add some extra loss $\eta_A^\prime$ to the channel between Alice and Charlie passively to satisfy $\mu_A\eta_A\eta_A^\prime=\mu_B\eta_B$ as in ~\cite{rubenok2013real}, the observed error rate will decrease, but the yield will decrease at the same time due to the higher loss.
The joint change of error rate and yield may not increase the final key rate a lot.
If we add extra loss $\frac{\mu_B\eta_B}{\mu_A\eta_A} \le \tilde{\eta}_A^\prime \le 1$ to the channel between Alice and Charlie, we can get a better key rate.
Given the transmittance and the intensities of sources, the specific value of $\tilde{\eta}_A^\prime$ can be determined by numerical simulation.

According to our numerical results, when $\mu_A\eta_A$ is close to $\mu_B\eta_B$, we don't need to add any extra loss ($\tilde{\eta}_A^\prime=1$) in the channel to get the best key rate.
Suppose that we are given the transmittance distribution $\{ \eta_A^{(1)} , \cdots , \eta_A^{(i)}, \cdots\}$ and $\{ \eta_B^{(1)} , \cdots , \eta_B^{(i)}, \cdots\}$ and fixed $\mu_A,\mu_B$.
When $\frac{\eta_A^{(i)}}{\eta_B^{(i)}}>\delta$ for a certain $\delta$, we should add some extra loss $\frac{\mu_B\eta_B}{\mu_A\eta_A} < \tilde{\eta}_A^\prime < 1$ to get the best best key rate.
In the case that $\mu_A\eta_A<\mu_B\eta_B$, we can add extra loss $\tilde{\eta}_B^\prime$ to the channel between Bob and Charlie similarly.

In the simulation of the unstable channel, suppose that we have the transmittance distribution $\{ \eta_A^{(1)} , \cdots , \eta_A^{(i)}, \cdots\}$, $\{ \eta_B^{(1)} , \cdots , \eta_B^{(j)}, \cdots\}$ and the corresponding probability $\{ p_A^{(1)} , \cdots , p_A^{(i)}, \cdots\}$, $\{ p_B^{(1)} , \cdots , p_B^{(j)}, \cdots\}$.
We can calculate the ``transmittance pair'' distribution
$\{ \eta_A^{(1)} \otimes  \eta_B^{(1)}, \cdots , \eta_A^{(i)} \otimes \eta_B^{(j)}, \cdots\}$ and the corresponding probability $\{ p_A^{(1)}*p_B^{(1)} , \cdots , p_A^{(i)}*p_B^{(j)}, \cdots\}$.
With a certain source pair $lr$ and a certain transmittance pair $\eta_A^{(i)} \otimes \eta_B^{(j)}$, the observed yield $S_{lr}(\eta_A^{(i)} \otimes \eta_B^{(j)})$ and the observed error rate $E_{lr}(\eta_A^{(i)} \otimes \eta_B^{(j)})$ can be calculated as in ~\cite{wang2014simulating} theoretically.
Then the yield and the error rate in the whole process can be calculated by
\begin{equation}\label{equ:Slr-com}
    S_{lr} = \sum_{i,j} p_A^{(i)} * p_B^{(j)} * S_{lr}(\eta_A^{(i)} \otimes \eta_B^{(j)})
\end{equation}
and
\begin{equation}\label{equ:Elr-com}
    E_{lr} = \sum_{i,j} p_A^{(i)} * p_B^{(j)} * E_{lr}(\eta_A^{(i)} \otimes \eta_B^{(j)}).
\end{equation}

When the loss-compensation is performed, we can calculate the $S_{lr},E_{lr}$ in the same way except the $(ij)$-th transmittance pair is changed into $\eta_A^{(i)}\tilde{\eta}^\prime \otimes \eta_B^{(j)}$ or $\eta_A^{(i)} \otimes \eta_B^{(j)}\tilde{\eta}^\prime$ if $\frac{\eta_A^{(i)}}{\eta_B^{(j)}}>\delta$ or $\frac{\eta_B^{(j)}}{\eta_A^{(i)}}>\delta$, respectively.

\section{Numerical simulation}\label{sec:numerical}
\begin{figure}[htb]
    \includegraphics[width=250pt]{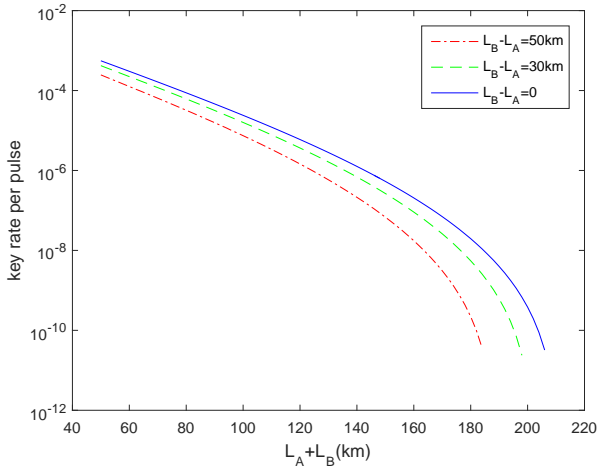}
    \caption{Optimized key rate versus the total distance between Alice and Bob with the device parameters in Table \ref{tab:device1}.}\label{fig:asym}
\end{figure}

First we consider the case that the channel is stable but asymmetric. We show the optimized key rate in some asymmetric cases in Fig.\ref{fig:asym} and some results in certain distances in Table~\ref{tab:asym-result} with device parameters in Table \ref{tab:device1}. From Fig.\ref{fig:asym}, we can find that with full implementation of the four-intensity MDIQKD and full optimization of the source parameters, the asymmetric channel doesn't decrease the key rate a lot at the same total distance.
\begin{table}[htb]
    \begin{tabular}{ccccccc}
    \hline
    $N_t$ &   $\eta_d$ &  $d$ &        $E_d^X$ &     $E_d^Z$ & $f$ & $\epsilon$\\
    \hline
    $10^{11}$ & $65\%$ & $8\times10^{-7}$ &   0.5\% &   0.5\% & 1.16 & $10^{-7}$ \\
    \hline
    \end{tabular}
    \caption{\label{tab:device1}Device parameters for Table \ref{tab:asym-result}. $N_t$: total number of pulse pairs; $\eta_d$: detection efficiency of the detectors; $d$: dark count rate of the detectors; $E_d^X$/$E_d^Z$: misalignment error rate in the $X$/$Z$ basis; $f$: correction efficiency; $\epsilon$: failure probability for statistical fluctuation evaluation between observable and the mean value.}
\end{table}

\begin{table}[htb]
    \begin{tabular}{|c|c|c|c|}
    \hline
    \multirow{2}{*}{$L_A$(km)} & \multirow{2}{*}{$L_B$(km)} & \multicolumn{2}{c|}{Optimized key rate per pulse pair}\\
    \cline{3-4}
     & & ours & Ref.\cite{wang2018enabling} \\
    \hline
    10  &  60 &    $6.299\times10^{-5}$ & $3.106\times10^{-5}$ \\
    \hline
    43  &  93  &   $3.151\times10^{-7}$ & $1\times10^{-8}$ \\
    \hline
    50  &  100  &   $6.576\times10^{-8}$ & $4.786\times10^{-11}$ \\
    \hline
    30  &  60  &   $3.117\times10^{-5}$ & $1.445\times10^{-5}$ \\
    \hline
    59.3  &  89.3  &   $2.972\times10^{-7}$ & $1\times10^{-8}$ \\
    \hline
    70 &  100 &   $2.490\times10^{-8}$ & 0 \\
    \hline
    \end{tabular}
    \caption{\label{tab:asym-result}Optimized key rate at different distances with the parameters in Table \ref{tab:device1}.}
\end{table}

Then for the unstable channel, we consider a simple case that $\eta_A^{(i)} = (3+2i)$dB, $\eta_B^{(j)} = (13+2j)$dB with probability $p_A^{(i)} = p_B^{(j)} = 0.2$ for $i,j=1,\cdots,5$.
We show the key rate with different $\delta$ and $\tilde\eta^\prime$ with source parameters optimized.
\begin{table}%[htb]
    \begin{tabular}{|c|c|c|}
    \hline
    $\delta$(dB) & $\tilde\eta^\prime$(dB) & Optimized key rate per pulse pair\\
    \hline
    0  &     0 &    $1.7747\times10^{-6}$   \\
    \hline
    -7  &  5  &   $1.5229\times10^{-6}$  \\
    \hline
    -8.5 &  4.5 &   $2.2279\times10^{-6}$  \\
    \hline
    -8.75 &  4 &   $2.2217\times10^{-6}$  \\
    \hline
    -8.75 &  4.5 &   $2.2283\times10^{-6}$  \\
    \hline
    -8.75 &  5 &   $2.2184\times10^{-6}$  \\
    \hline
    -9 &  4.5 &   $2.2278\times10^{-6}$  \\
    \hline
    \end{tabular}
    \caption{\label{tab:unstable-result}Optimized key rate with different $\delta$ and $\tilde\eta^\prime$ with the parameters in Table \ref{tab:device1}.}
\end{table}
The case that $\delta=\tilde\eta^\prime=0$dB means that we don't perform the loss-compensation method. In the case that $\delta=-7$dB and $\tilde\eta^\prime=5$dB, we can see that an improper loss-compensation will decrease the key rate. We can find that in this transmittance distribution, setting $\delta=-8.75$dB and $\tilde\eta^\prime=4.5$dB can maximize the key rate in our loss-compensation method.

\section{Concluding remark}\label{sec:Conclusion}
We propose a full implementation of four-intensity decoy-state MDIQKD. Even if the channels between the users and UTP are asymmetric, our four-intensity protocol still has a good performance. We also propose a loss-compensation method. This method can improve the key rate a lot in unstable channel.

{\bf Acknowledgement} We acknowledge the financial support in part by
The National Key Research and Development Program of China grant No. 2017YFA0303901; NSFC grant No. 11474182, 11774198 and U1738142; the key Research and Development Plan Project of Shandong Province,  grant No. 2015GGX101035; Shandong Peninsula National Innovation Park Development Project; Taishan Scholars of Shandong Province. Yuan Cao was supported by the Youth Innovation Promotion Association of CAS.

Yuan Cao and Xiao-Long Hu have equally contributed to the work.

%the 10000-Plan of Shandong province (Taishan Scholars), National High-Tech Program of China grant No. 2011AA010800 and 2011AA010803,
%NSFC grant No. 11474182, 11174177 and 60725416, the key R$\&$D Plan Project of Shandong Province,  grant No. 2015GGX101035, and Open Research Fund Program of the State Key Laboratory of Low-Dimensional Quantum Physics Grant No. KF201513.

%\clearpage
%\bibliographystyle{unsrt}
%\bibliography{refs}

\end{document}